\documentclass[prc,twocolumn,superscriptaddress,showpacs,psfig,showkeys,nofootinbib]{revtex4}
\usepackage{amsmath}
\usepackage{amssymb}
\usepackage{amsfonts}
\usepackage{graphicx}
\usepackage{epsfig}
\usepackage{dcolumn}
\usepackage{hyperref}
\usepackage{epstopdf}
\usepackage{color}
\usepackage[utf8]{inputenc}
\usepackage{amsthm}
\usepackage{fontenc}
\usepackage{bm}
\RequirePackage{slashed}\usepackage{cancel}
\usepackage[normalem]{ulem}

\usepackage{array,booktabs}
\newcolumntype{C}{>{$\displaystyle}c<{$}}

\usepackage{soul}

\newcommand \Pomeron {I\!\!P}

\begin{document}


\title{Contributions from ultraperipheral collisions to the forward rapidity gap distribution in $p$Pb collisions at the CERN Large Hadron Collider at $\sqrt{s_{NN}}=8.16$ TeV}

\author{V. Guzey}
\affiliation{National Research Center ``Kurchatov Institute'', Petersburg Nuclear Physics Institute (PNPI), Gatchina, 188300, Russia}
\author{M. Strikman}
\affiliation{Pennsylvania  State  University,  University  Park,  PA,  16802,  USA} 
\author{M. Zhalov}
\affiliation{National Research Center ``Kurchatov Institute'', Petersburg Nuclear Physics Institute (PNPI), Gatchina, 188300, Russia}

\date{\today}

\begin{abstract}

In this letter, we consider strong and electromagnetic (ultraperipheral) mechanisms in proton-nucleus coherent diffraction at the LHC. We explicitly demonstrate the dominance of the latter and explain the CMS data on the forward rapidity gap distribution 
in $pPb$ collisions at $\sqrt{s_{NN}}=8.16$ TeV. In particular, we provide simple estimates, which give a good, 
semi-quantitative description of both magnitude and shape of the $\Delta \eta^F$ distribution in the Pomeron-proton topology.
We also make predictions for the proton-oxygen run.

\end{abstract}

\maketitle

\textbf{\textit{Introduction and motivation}}.
Diffraction in hadron scattering at high energies remains an active field of research. It is deeply connected to the nature 
of colorless exchanges with vacuum quantum numbers (Pomeron) in strong interactions and, more  generally, small-$x$ phenomena in Quantum Chromodynamics 
(QCD), important for tuning event generators needed for interpretation of results of ultrarelativistic heavy-ion scattering, and  
 also relevant for cosmic ray physics. In experiment diffractive events are characterized by large gaps in 
 rapidity distributions of produced particles, which are defined as regions with no hadronic activity. 
 To enhance sensitivity to such events and, in particular, to the so-called single diffractive dissociation, 
 one can select events with the rapidity gaps in the most forward region of a detector; in proton-proton ($pp$) scattering such measurements have been performed at the Large Hadron Collider (LHC) at $\sqrt{s_{NN}}=7$ TeV~\cite{ATLAS:2012djz,CMS:2015inp}.

The CMS collaboration at the LHC for the first time measured the forward rapidity gap distribution in proton-Pb ($pPb$) collisions at $\sqrt{s_{NN}}=8.16$ TeV~\cite{Sosnov:2022abz}. It was found that for the Pomeron-proton topology, the EPOS-LHC,  QGSJET II, and HIJUNG generators are at least a factor of five below the data. As a result, it was suggested that this 
discrepancy can be explained by a significant 
contribution of ultraperipheral photoproduction events mimicking the signature of diffractive processes.

Actually, this observation was already made in Ref.~\cite{Guzey:2005tk} in 2006, which showed that in coherent proton-nucleus ($pA$) diffraction, the electromagnetic (ultraperipheral) contribution dominates the cross section for heavy nuclei.
The purpose of this letter is to generalize the results of~\cite{Guzey:2005tk} 
to the CMS experimental conditions and, in particular, to make predictions for the distribution in the forward rapidity gap $\Delta \eta^F$. 
Our predictions for the 
$\Delta \eta^F$ distribution in the studied case of the Pomeron-proton topology agree both in magnitude and the shape with that measured by the CMS collaboration and, thus, 
confirm and quantify the essential role of ultraperipheral photoproduction in explanation of the CMS data.

We also make predictions for the case of proton-oxygen ($pO$) scattering.

\textbf{\textit{Strong and electromagnetic mechanisms in $pA$ coherent diffraction}}.
The phenomenon of diffractive dissociation of protons in proton-nucleus scattering at high energies is a classic example of 
composite structure of hadronic projectiles, which can be conveniently described within the framework of cross section
fluctuations~\cite{Good:1960ba,Frankfurt:1993qi,Blaettel:1993ah,Frankfurt:2000tya,Frankfurt:2022jns}. In this approach,
the cross section of $pA$ coherent diffraction dissociation, $p+A \to X+A$, can be written in the following form
\begin{eqnarray}
&&\sigma_{pA}^{\rm diff}(s)= \nonumber\\
&& \int d^2{\vec b} \left[\int d\sigma P_p(\sigma)|\Gamma_A({\vec b})|^2
-\left|\int d\sigma P_p(\sigma) \Gamma_A({\vec b})\right|^2 \right] \,,
\label{eq:sigma_dd}
\end{eqnarray}
where $s$ is the total proton-nucleus energy squared per nucleon.
Here $\Gamma_A({\vec b})$ is the nuclear scattering amplitude in representation of the impact parameter ${\vec b}$, which in
the limit of high energies and large $A$ (heavy nucleus) is usually expressed in the eikonal form 
\begin{equation}
\Gamma_A({\vec b})=1-e^{-\frac{\sigma}{2}T_A({\vec b})} \,,
\label{eq:Gamma_A}
\end{equation}
where $T_A(\vec{b})=\int dz \rho_A({\vec r})$ with $\rho_A({\vec r})$ being the nuclear density~\cite{DeVries:1987atn} 
normalized to the number of nucleons $A$. 
The  $\Gamma_A({\vec b})$ amplitude sums multiple interactions with target nucleons and captures the effect of nuclear shadowing
leading to a dramatic suppression of the proton-nucleus cross section.

The distribution $P_p(\sigma)$ describes cross section fluctuations of the proton and gives the probability for the proton to
fluctuate into a hadronic configuration interacting with target nucleons with the cross section $\sigma$.
In general, $P_p(\sigma)$ should be modeled, see, e.g.~\cite{Blaettel:1993ah,Frankfurt:2022jns}. 
However, in the case of diffraction dissociation, the detailed information on the shape of $P_p(\sigma)$ is not needed since one can use the general property that $P_p(\sigma)$ is peaked around 
$\sigma_{pp}^{\rm tot}(s)=\langle \sigma \rangle \equiv \int d\sigma P_p(\sigma) \sigma$.
Thus, expanding Eq.~(\ref{eq:sigma_dd}) around $\langle \sigma \rangle$, one obtains~\cite{Frankfurt:1993qi}
\begin{equation}
\sigma_{pA}^{\rm diff}(s)=\frac{\omega_{\sigma}(s) \langle \sigma \rangle^2}{4}
\int d^2{\vec b} \left(T_A(\vec{b})\right)^2 e^{-\langle \sigma \rangle T_A(\vec{b})}  \,,
\label{eq:sigma_dd2}
\end{equation}
where $\omega_{\sigma}(s)=\langle \sigma^2 \rangle/\langle \sigma \rangle^2-1$ quantifies the dispersion of cross section fluctuations of the proton.
At $\sqrt{s}=\sqrt{s_{NN}}=8.16$ TeV, we use the COMPETE parametrization~\cite{ParticleDataGroup:2020ssz} giving $\langle \sigma \rangle = \sigma_{pp}^{\rm tot}(s)= 98.6$ mb and a simple interpolation from fixed-target to Tevatron and further 
extrapolation to LHC energies giving 
$\omega_{\sigma}(s)=0.092 \pm 0.050$~\cite{Frankfurt:2022jns}. The spread in the values of $\omega_{\sigma}(s)$ reflects 
the theoretical uncertainty in modeling $P_p(\sigma)$.
Note that the accuracy of approximating Eq.~(\ref{eq:sigma_dd}) by Eq.~(\ref{eq:sigma_dd2}) is better than the significant uncertainty of $\omega_{\sigma}(s)$, which is amplified by its extrapolation from Tevatron to LHC energies and which dominates 
the uncertainty of the predicted values of $\sigma_{pA}^{\rm diff}(s)$.

It was explained in~\cite{Guzey:2005tk} that a competing reaction mechanism leading to the same final state, 
$p+A \to p+\gamma+A \to X+A$,
is provided by the electromagnetic contribution corresponding to ultraperipheral $pA$ scattering. In this case, proton
and Pb beams pass each other at large impact parameters and, hence, short-range strong interactions are suppressed. Instead, 
 the relativistic heavy ion beam serves as an intensive source of
quasi-real photons, which interact with the proton. In the equivalent photon (Weizs\"acker-Williams) approximation, 
the corresponding cross section reads~\cite{Baur:2001jj,Baltz:2007kq}
\begin{equation}
\sigma_{pA}^{\rm e.m.}(s) =\int^{\omega_{\rm max}}_{\omega_{\rm min}} \frac{d \omega}{\omega} N_{\gamma/A}(\omega)\sigma_{\gamma p}^{\rm tot}(s_{\gamma p}) \,,
\label{eq:sigma_em}
\end{equation}
where $N_{\gamma/A}(\omega)$ is the photon flux; 
$\omega$ is the photon energy; $\sigma_{\gamma p}^{\rm tot}(s_{\gamma p})$ is the total photon-proton cross section
and $s_{\gamma p}$ is the total invariant photon-proton energy squared.
The integration limits can be estimated as follows. In the laboratory frame, the minimal photon energy corresponding 
to photo-excitation of the lowest inelastic state is $\omega_{\rm min}=(M_{\Delta}^2-m_p^2)/(4 m_p \gamma_L(p))$, 
where $M_{\Delta}$ and $m_p$ are the masses of $\Delta(1232)$ and the proton, respectively, and 
$\gamma_L(p)=E_p/m_p$ is the Lorentz factor of the proton beam with energy $E_p$. The maximal photon energy
is usually estimated as $\omega_{\rm max} = \gamma_L(A)/R_A$, where $R_A$ is the nucleus effective radius and 
$\gamma_L(A)=E_A/m_p$ is the Lorentz factor of the nucleus beam with energy $E_A$.

For the photon flux, we use the approximate expression corresponding to the point-like (PL) source with the electric 
charge $Z$:
\begin{eqnarray}
&& N_{\gamma/A}(\omega)= \nonumber\\
&&\frac{2 Z^2 \alpha_{\rm e.m.}}{\pi} \left(\xi K_0(\xi)K_1(\xi)-\frac{\xi^2}{2} (K_1^2(\xi)-K_0^2(\xi)) 
\right) \,,
\label{eq:flux}
\end{eqnarray}
where $\alpha_{\rm e.m.}$ is the fine-structure constant; $K_{0,1}$ are modified Bessel functions of the 
second kind; 
 $\xi=(\omega/\gamma_L(A)) b_{\rm min}$ with $b_{\rm min}=1.15 R_A$ and $R_A=1.145 A^{1/3}$ fm.
 With these parameters, Eq.~(\ref{eq:flux}) reproduces 
with a 5\% precision
  a more accurate calculation of the photon flux 
 taking into account the suppression of strong interactions at $|{\vec b}| \leq b_{\rm min}$~\cite{Guzey:2013taa}.
 This estimate of the accuracy of Eq.~(\ref{eq:flux}) is based on the analysis of Ref.~\cite{Guzey:2013taa}
 and also includes the effect of the use of different nuclear density distributions.

For the total photon-proton cross section, we use the Donnachie and Landshoff fit~\cite{Donnachie:1992ny}
\begin{equation}
\sigma_{\gamma p}^{\rm tot}(s)/{\rm mb}=0.0677 s_{\gamma p}^{0.0808}+0.129 s_{\gamma p}^{-0.4525} \,,
\label{eq:DL}
\end{equation}
where $s_{\gamma p}=4 \omega E_p+m_p^2$.

Employing the input specified above and using Eqs.~(\ref{eq:sigma_dd2}) and (\ref{eq:sigma_em}), we obtain the
following results for the strong and electromagnetic (ultraperipheral) contributions to the cross section of $pPb$ 
coherent diffraction at $\sqrt{s_{NN}}=8.16$ TeV
\begin{eqnarray}
\sigma_{pA}^{\rm diff}(s) &=& 7.4 \pm 4.0 \ {\rm mb} \,, \nonumber\\
\sigma_{pA}^{\rm e.m.}(s) &=& 450 \pm 23 \ {\rm mb} \,.
\label{eq:res1}
\end{eqnarray} 
These values agree with those of Ref.~\cite{Guzey:2005tk} (the correct predictions for the electromagnetic 
contribution are given in the Erratum to that paper) and of Ref.~\cite{Goncalves:2019agu}.
The uncertainty in the predicted value of $\sigma_{pA}^{\rm e.m.}(s)$ comes from the uncertainty in 
$N_{\gamma/A}(\omega)$.

\begin{figure}[h]
\begin{center}
\epsfig{file=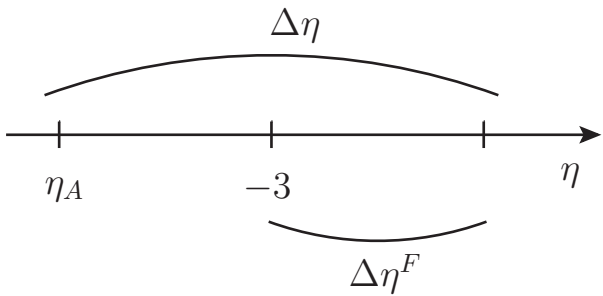,scale=1.0}
\caption{Sketch of the definition of the rapidity gap size $\Delta \eta^F$ in the Pomeron-proton topology at CMS.}
\label{fig:gap}
\end{center}
\end{figure}

\textbf{\textit{Predictions for the strong and electromagnetic contributions differential in $\Delta \eta^F$}}.
In proton-nucleus coherent diffraction, the size of the rapidity gap between the intact nucleus 
and the diffractively-produced system $X$ is
\begin{equation}
\Delta \eta=-\ln \xi_X \,,
\label{eq:Delta_eta}
\end{equation}
where $\xi_X=M_X^2/s$ is a variable commonly used in diffraction and $M_X$ is the mass of the state $X$.
In the case of Pomeron-proton topology, 
the CMS collaboration has defined $\Delta \eta^F$ as the distance from $\eta=-3$ to the lower edge of the last 
non-empty $\eta$ bin~\cite{Sosnov:2022abz}. Since the elastically scattered nucleus corresponds to
$\eta_A=-(1/2) \ln (4 E_A^2/m_p^2)=\ln (2 E_A/m_p)=-8.6$ (in the CMS coordinate system, the direction of the proton beam 
in $pPb$ collisions defines positive rapidity), we obtain
\begin{equation}
\Delta \eta^F=\Delta \eta-(8.6-3)=\Delta \eta-5.6 \,.
\label{eq:Delta_eta_CMS}
\end{equation}
This is illustrated in Fig.~\ref{fig:gap}.
It should be compared to the definition of the ATLAS collaboration in the $pp$ case at $\sqrt{s_{NN}}=7$ TeV,
$\Delta \eta^F= \Delta \eta-4$~\cite{ATLAS:2012djz}.

Turning to Eq.~(\ref{eq:sigma_dd2}) and recalling that the cross section of diffraction dissociation on the proton (nucleon)
at the momentum transfer $t=0$
is related to the dispersion of cross section fluctuations~\cite{Good:1960ba},
\begin{equation}
\frac{d\sigma_{pp}^{\rm diff}(t=0)}{dt}=\frac{1}{16 \pi} \left(\langle \sigma^2 \rangle -\langle \sigma \rangle^2 \right)
=\frac{\omega_{\sigma}(s) \langle \sigma \rangle^2}{16 \pi} \,,
\label{eq:sigma_dd_p}
\end{equation}
Eq.~(\ref{eq:sigma_dd2}) can be rewritten in the following form
\begin{equation}
\sigma_{pA}^{\rm diff}(s)= \frac{d\sigma_{pp}^{\rm diff}(t=0)}{dt}
4 \pi \int d^2{\vec b} \left(T_A(b)\right)^2 e^{-\langle \sigma \rangle T_A(b)}  \,.
\label{eq:sigma_dd3}
\end{equation}
Making the common assumption of an exponential momentum transfer $t$ dependence, 
$d\sigma_{pp}^{\rm diff}/dt=e^{-B(s) |t|}d\sigma_{pp}^{\rm diff}(t=0)/dt$,
we can express the proton-nucleus diffractive cross section as a product of the $t$-integrated proton-proton
diffractive cross section $\sigma_{pp}^{\rm diff}(s)$ and the nuclear factor, 
\begin{eqnarray}
\sigma_{pA}^{\rm diff}(s)& = &  \sigma_{pp}^{\rm diff}(s)
4 \pi B(s) \int d^2{\vec b} \left(T_A(b)\right)^2 e^{-\langle \sigma \rangle T_A(b)} \nonumber\\
&=& (2.4 \pm 0.16) \, \sigma_{pp}^{\rm diff}(s) \,.
\label{eq:sigma_dd4}
\end{eqnarray}
In the second line of Eq.~(\ref{eq:sigma_dd4}), we used that $B(s) \approx B_{\rm el}+2 \alpha_{\Pomeron}^{\prime} \ln (m_p^2/M_X^2) 
\approx 15 \pm 1$ GeV$^{-2}$ for $40 \leq M_X \leq 300$ GeV at
$\sqrt{s_{NN}}=8.16$ TeV.
This estimate is based on the experimental results for the slope of the $t$ dependence of the elastic $pp$ cross section
$B_{\rm el} \approx 20 \pm 0.5$ GeV$^{-2}$~\cite{ParticleDataGroup:2020ssz} and the general dependence of the slope of single diffractive dissociation on $M_X^2$ in Regge phenomenology with $\alpha_{\Pomeron}^{\prime} \approx 0.25$ GeV$^{-2}$;
the used range of $M_X$ corresponds to $1 \leq \Delta \eta^F \leq 5$.

Equation~(\ref{eq:sigma_dd4}) has a transparent probabilistic interpretation: 
the process of diffractive dissociation with the cross section 
$d\sigma_{pp}^{\rm diff}(t=0)/dt=B(s) \sigma_{pp}^{\rm diff}(s)$ takes place coherently on nuclear target nucleons, 
which are distributed in the transverse plane with the probability $(T_A(b))^2$; 
the probability to maintain nuclear coherence, i.e., the probability not to have inelastic interactions within the nuclear volume, 
is given by the standard Glauber model factor $e^{-\langle \sigma \rangle T_A(b)}$.

This factorization (decoupling) of diffractive dissociation on the nucleon from the effect of nuclear suppression, which 
does not depend on the diffraction,
 allows for a simple generalization to the case of cross sections differential in the produced diffractive mass
$M_X$ (the variable $\xi_X)$ or the size of the rapidity gap $\Delta \eta^F$, see Eqs.~(\ref{eq:Delta_eta})
and (\ref{eq:Delta_eta_CMS}). Indeed,
taking advantage of a simple connection between $\sigma_{pA}^{\rm diff}(s)$ and $\sigma_{pp}^{\rm diff}(s)$
and neglecting a weak dependence on $\xi_X$ of the slope $B(s)$ and the nuclear factor in Eq.~(\ref{eq:sigma_dd4}), we
can generalize Eq.~(\ref{eq:sigma_dd4}) to the form differential in $\Delta \eta^F$, 
\begin{equation}
\frac{d \sigma_{pA}^{\rm diff}}{d \Delta \eta^F}=(2.4 \pm 0.16) \,\frac{d\sigma_{pp}^{\rm diff}}{d \Delta \eta^F} \,.
\label{eq:sigma_dd5}
\end{equation}
Finally, without resorting to a particular model for $d\sigma_{pp}^{\rm diff}/d \Delta \eta^F$, we use the
ATLAS result
that $d\sigma_{pp}^{\rm diff}/d \Delta \eta^F \approx 1 \ {\rm mb}$ 
for $\Delta \eta^F \geq 3$~\cite{ATLAS:2012djz} and thus arrive at the following estimate,
\begin{equation}
\frac{d \sigma_{pA}^{\rm diff}}{d \Delta \eta^F} \approx 2.4 \pm 1.3  \ {\rm mb} \,.
\label{eq:sigma_dd6}
\end{equation}
The uncertainty in this estimate comes from the significant uncertainty in the value of $\omega_{\sigma}(s)$ discussed above, 
see the first line of Eq.~(\ref{eq:res1}).
Note that Eq.~(\ref{eq:sigma_dd6}) becomes less accurate for small values of the rapidity gap
since in the $pp$ case for $\Delta \eta^F < 2$, non-diffractive processes dominate and $d\sigma_{pp}^{\rm diff}/d \Delta \eta^F$ begins 
to rapidly grow
(see the CMS data point at $\Delta \eta^F < 1$ in Fig.~\ref{fig:res} below).

The estimate of Eq.~(\ref{eq:sigma_dd6}) semi-quantitatively agrees with predictions of EPOS-LHC,  QGSJET II, and HIJUNG generators
shown in Fig.~2 of Ref.~\cite{Sosnov:2022abz}.
In particular, for $\Delta \eta^F \geq 2$, it agrees on a logarithmic scale with the approximately constant predictions of QGSJET II, 
$d\sigma/d \Delta \eta^F \approx 2$ mb, and of EPOS-LHC and HIJING, $d\sigma/d \Delta \eta^F \approx 4$ mb.
For $\Delta \eta^F < 2$, results of these event generators tend to somewhat increase.

Turning to Eq.~(\ref{eq:sigma_em}), we notice that the photon energy required to excited the diffractive mass $M_X$ is 
$\omega=(M_X^2-m_p^2)/(4 m_p \gamma_L(p))  \approx M_X^2/(4 m_p \gamma_L(p))$ for sufficiently large $M_X$.
Therefore,
\begin{equation}
\frac{d\omega}{\omega}= d \ln M_X^2=d \Delta \eta^F \,.
\label{eq:trick}
\end{equation}
It allows us to rewrite Eq.~(\ref{eq:sigma_em}) in the form differential in $\Delta \eta^F$
\begin{equation}
\frac{d\sigma_{pA}^{\rm e.m.}}{d \Delta \eta^F} = N_{\gamma/A}(\omega(\Delta \eta^F))\sigma_{\gamma p}^{\rm tot}(s_{\gamma p}) \,,
\label{eq:sigma_em2}
\end{equation}
where the photon energy corresponds to the given $\Delta \eta^F$, i.e., to the given $M_X$, see Eqs.~(\ref{eq:Delta_eta}) and (\ref{eq:Delta_eta_CMS}). The resulting values of $d\sigma_{pA}^{\rm e.m.}/d \Delta \eta^F$ as a function of $\Delta \eta^F$ in the $1 \leq \Delta \eta^F \leq 5$ interval are given in Table~\ref{table:res} (second column).
The 5\% uncertainty comes from the uncertainty in the photon flux $N_{\gamma/A}(\omega)$.
In the last column of the table, we give a sum of the electromagnetic and strong interaction contributions (total cross section),
where the respective uncertainties have been added in quadrature.

\begin{table}[h]
\caption{The contribution of the electromagnetic (ultraperipheral) mechanism to $pPb$ coherent diffraction,
$d\sigma_{pA}^{\rm e.m.}/d \Delta \eta^F$, and a sum of the electromagnetic and strong interaction contributions 
(total cross section), $d\sigma_{pA}/d \Delta \eta^F$,
 as a function of the rapidity gap size $\Delta \eta^F$.}
\begin{center}
\begin{tabular}{|c|c|c|}
\hline
$\Delta \eta^F$ & $d\sigma_{pA}^{\rm e.m.}/d \Delta \eta^F$, mb & $d\sigma_{pA}/d \Delta \eta^F$, mb\\
\hline
1 & $13.9 \pm  0.70$ & $16.3  \pm 1.48 $ \\
2 & $17.8  \pm 0.89$ & $20.2  \pm 1.58 $\\
3 &  $21.1 \pm 1.06$ & $23.5  \pm 1.68 $\\
4 &  $23.9 \pm 1.20$ & $26.3  \pm 1.77 $\\
5 &  $26.3 \pm 1.32$ & $28.7  \pm 1.85 $\\
\hline
\end{tabular}
\end{center}
\label{table:res}
\end{table}%

In a graphical form
our results are summarized in Fig.~\ref{fig:res}. It shows the strong (green dot-dashed curve labeled ``diff''), 
electromagnetic (blue dotted curve labeled ``e.m.''), and total (the sum of the former two given by the red solid curve labeled ``Total'') contributions
to the cross section of proton-lead ($pPb$) coherent diffraction as a function of $\Delta \eta^F$.
The shaded bands represent uncertainties of our predictions detailed above; the band for the total cross section is obtained by adding
in quadrature the uncertainties of the strong and electromagnetic contributions.
The preliminary CMS data are shown by solid circles with error bars; we extracted them from~\cite{Sosnov:2022abz} using the WebPlotDigitizer tool~\cite{WebPlot}. 
\begin{figure}[h]
\begin{center}
\epsfig{file=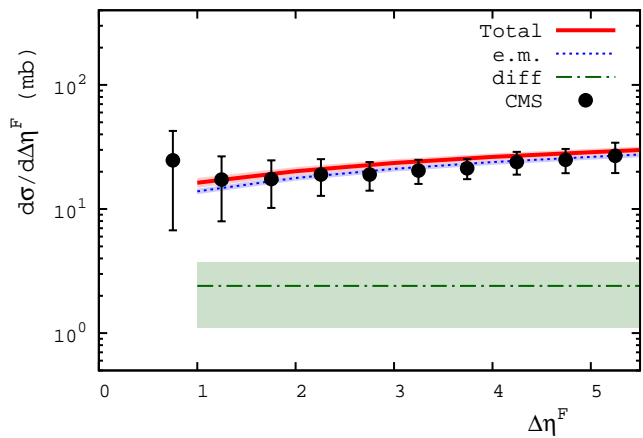,scale=1.0}
\caption{The strong (``diff''), electromagnetic (``e.m.''), and total (``Total'') contributions
to the cross section of $pPb$ coherent diffraction as a function of $\Delta \eta^F$ at $\sqrt{s_{NN}}=8.16$ TeV. 
The preliminary CMS data~\cite{Sosnov:2022abz} are shown by solid circles with error bars. }
\label{fig:res}
\end{center}
\end{figure}

A comparison to the CMS results presented in Fig.~\ref{fig:res}
shows that our simple estimate reproduces rather
well both the magnitude and the shape of the measured $\Delta \eta^F$ distribution, 
which exhibits a slow, monotonous increase from $d\sigma/d \Delta \eta^F \approx 20$ mb at $\Delta \eta^F=1$ to 
$d\sigma/d \Delta \eta^F \approx 30$ mb at $\Delta \eta^F=5$.
Thus, we demonstrate that the ultraperipheral mechanism 
is responsible for the increase of $d\sigma/d \Delta \eta^F$ with an increase of $\Delta \eta^F$.
Note that the $\Delta \eta^F < 1$ region,  where $d\sigma_{pp}^{\rm diff}/d \Delta \eta^F$ begins 
to rapidly grow, corresponds non-diffractive processes and, hence, is outside of the range of applicability of our approach.

It is important to note that our estimate of $d\sigma_{pA}^{\rm e.m.}/d \Delta \eta^F$ is based on the assumption that 
it receives contributions from all $M_X$ comprising the total photon-proton cross section and, hence, 
should be considered as an upper limit.
A more accurate account of the
ultraperipheral contribution to $d\sigma/d \Delta \eta^F$ should include modeling of the mass spectrum in photon-proton scattering and the influence of the detector acceptance, which is beyond the scope of our work.

While the aim of this letter was to capture the bulk of physical effects explaining the CMS results in a semi-quantitative way, 
our calculations can be improved along several lines, in particular, in an estimate of the strong interaction mechanism of coherent diffraction. However, since it gives a subleading contribution, these refinements will not significantly affect the resulting total 
$\Delta \eta^F$ distribution.  

\textbf{\textit{Predictions for proton-oxygen run}}.
One can readily extend our predictions to proton-oxygen ($pO$) scattering at $\sqrt{s_{NN}}=9.19$ TeV. In this case, 
$\sigma_{pp}^{\rm tot}(s)= 100.6$ mb and $\omega_{\sigma}(s)=0.086 \pm 0.050$, and we obtain (compare to Eq.~(\ref{eq:res1}))
\begin{eqnarray}
\sigma_{pO}^{\rm diff}(s) &=& 3.1 \pm 1.8 \ {\rm mb} \,, \nonumber\\
\sigma_{pO}^{\rm e.m.}(s) &=& 5.0 \pm 0.25\ {\rm mb} \,.
\label{eq:res1_O}
\end{eqnarray} 
One can see that the strong interaction and electromagnetic contributions have comparable magnitudes for oxygen because of 
a 100 times smaller photon flux compared to Pb. As a result, the electromagnetic contribution constitutes a $15-30$\% correction
to the $\Delta \eta^F$ distribution.
At the same time, this gives an opportunity to measure the cross section of soft $pO$ diffraction,
which is strongly suppressed by nuclear shadowing compared to the impulse approximation.

\textbf{\textit{Summary}}.
In summary, we showed that a straightforward extension of the results of Ref.~\cite{Guzey:2005tk} 
can explain the CMS data on the forward rapidity gap distribution in $pPb$ collisions 
at $\sqrt{s_{NN}}=8.16$ TeV. Notably, we explicitly demonstrated the dominance of the electromagnetic (ultraperipheral)
mechanism in the Pomeron-proton topology, which provides a good, semi-quantitative description of both magnitude and shape of 
the measured
$\Delta \eta^F$ distribution.

{\textbf{\textit{Acknowledgements}}.
The research of M.S was supported by the US Department of Energy Office of Science, Office of Nuclear Physics under 
Award No.~DE-FG02-93ER40771.
MS thanks Theory Division of CERN for hospitality while this work was done.


\begin{thebibliography}{99}

\bibitem{ATLAS:2012djz}
G.~Aad \textit{et al.} [ATLAS],
Eur. Phys. J. C \textbf{72}, 1926 (2012)
[arXiv:1201.2808 [hep-ex]].

\bibitem{CMS:2015inp}
V.~Khachatryan \textit{et al.} [CMS],
Phys. Rev. D \textbf{92}, no.1, 012003 (2015)
[arXiv:1503.08689 [hep-ex]].

\bibitem{Sosnov:2022abz}
D.~Sosnov [CMS],
Phys. Part. Nucl. \textbf{53}, no.2, 393-397 (2022);
%
 [CMS],
CMS-PAS-HIN-18-019.

\bibitem{Guzey:2005tk}
V.~Guzey and M.~Strikman,
Phys. Lett. B \textbf{633}, 245-252 (2006),
Phys. Lett.B \textbf{663} 456 (2008) 
[arXiv:hep-ph/0505088 [hep-ph]]; 

\bibitem{Good:1960ba}
M.~L.~Good and W.~D.~Walker,
Phys. Rev. \textbf{120}, 1857-1860 (1960)

\bibitem{Frankfurt:1993qi}
L.~Frankfurt, G.~A.~Miller and M.~Strikman,
Phys. Rev. Lett. \textbf{71}, 2859-2862 (1993)
[arXiv:hep-ph/9309285 [hep-ph]].

\bibitem{Blaettel:1993ah}
B.~Blaettel, G.~Baym, L.~L.~Frankfurt, H.~Heiselberg and M.~Strikman,
Phys. Rev. D \textbf{47}, 2761-2772 (1993)

\bibitem{Frankfurt:2000tya}
L.~Frankfurt, V.~Guzey and M.~Strikman,
J. Phys. G \textbf{27}, R23-146 (2001)
[arXiv:hep-ph/0010248 [hep-ph]].

\bibitem{Frankfurt:2022jns}
L.~Frankfurt, V.~Guzey, A.~Stasto and M.~Strikman,
[arXiv:2203.12289 [hep-ph]].

\bibitem{DeVries:1987atn}
H.~De Vries, C.~W.~De Jager and C.~De Vries,
Atom. Data Nucl. Data Tabl. \textbf{36}, 495-536 (1987)

\bibitem{ParticleDataGroup:2020ssz}
P.~A.~Zyla \textit{et al.} [Particle Data Group],
PTEP \textbf{2020}, no.8, 083C01 (2020)

\bibitem{Baur:2001jj}
G.~Baur, K.~Hencken, D.~Trautmann, S.~Sadovsky and Y.~Kharlov,
Phys. Rept. \textbf{364}, 359-450 (2002)
[arXiv:hep-ph/0112211 [hep-ph]].

\bibitem{Baltz:2007kq}
A.~J.~Baltz, G.~Baur, D.~d'Enterria, L.~Frankfurt, F.~Gelis, V.~Guzey, K.~Hencken, Y.~Kharlov, M.~Klasen and S.~R.~Klein, \textit{et al.}
Phys. Rept. \textbf{458}, 1-171 (2008)
[arXiv:0706.3356 [nucl-ex]].

\bibitem{Guzey:2013taa}
V.~Guzey and M.~Zhalov,
JHEP \textbf{02}, 046 (2014)
[arXiv:1307.6689 [hep-ph]].

\bibitem{Donnachie:1992ny}
A.~Donnachie and P.~V.~Landshoff,
Phys. Lett. B \textbf{296}, 227-232 (1992)
[arXiv:hep-ph/9209205 [hep-ph]].

\bibitem{Goncalves:2019agu}
V.~P.~Gon\c{c}alves, R.~P.~da Silva and P.~V.~R.~G.~Silva,
Phys. Rev. D \textbf{100}, no.1, 014019 (2019)
[arXiv:1905.00806 [hep-ph]].

\bibitem{WebPlot}
{\tt https://apps.automeris.io}


\end{thebibliography}
\end{document}